# Nanotechnology Applications

# The future arrived suddenly

Manuel Alberto M. Ferreira

Instituto Universitário de Lisboa (ISCTE-IUL), Business Research Unit (BRU-IUL) and Information Sciences, Technologies and Architecture Research Center (ISTAR-IUL), **PORTUGAL**

José António Filipe

Instituto Universitário de Lisboa (ISCTE-IUL), Business Research Unit (BRU-IUL) and Information Sciences, Technologies and Architecture Research Center (ISTAR-IUL), **PORTUGAL**

E-mails : manuel.ferreira@iscte.pt

jose.filipe@iscte.pt

**ABSTRACT**

*There is already a significant time, but it gives the sensation of extremely short, nanotechnology has become one of the most promising scientific hopes in innumerable human domains. Now the hope become reality. Countless scientific studies in several areas of knowledge have been made since the nanoscale emergence, carrying their contribution to the nanoscience development. The recent research in this field allowed the union of interests among several areas, such as physical sciences, molecular engineering, biology, biotechnology and medicine for example, contributing to the investigation of biosystems at a nanoscale. In this work begin discussing nanotechnology in a general way. Then nanotechnology and the applications in industry, in electronics and in medicine are presented and some discussion is proposed in order to define the boundaries for the advances on those areas. In the end, nanotechnology is discussed in terms of ethics and in terms of the borders that nanotechnology applications must satisfy and concluding notes are presented, highlighting the results of the analysis. Important considerations are made about the close connection between ethics and the*

*nanotechnology and the effects over the society and values. Some future directions for the research are suggested.*

**Keywords:** *Nanotechnology, nanomaterials, nanoelectronics, telemedicine, ethics.*

## INTRODUCTION

Beginning in mili ($10^{-3}$), then micro ($10^{-6}$) and now nano ($10^{-9}$). The "small" is becoming smaller and smaller. This is evidence in a lot of domains and with great impact in our lives as it happens with this technology, which maybe paradoxically has the greatest one. In short, "small is beautiful". Briefly, of course, pico ($10^{-12}$), then femto ($10^{-15}$) will arrive. However, for now, stand in nano world.

Nanotechnology - extremely small technology at the nanometer scale - involves multidisciplinary scientific knowledge and discusses matters related to the use of materials, mechanisms, devices and systems at a nanometer scale. Nowadays there are a lot of fields in which nanotechnology is applied and it is expected that it can create an enormous field for innovations and to be part in many fields of science developments. Particularly, it is expected that these developments happen in various applications on the biomedical area, such as drug delivery, molecular imaging, biomarkers, biosensors (biomedical sensors wirelessly located on/in the human body; nanobiomedical data transmitter) and many other fields of application.

As far as it is known, the first scientist to conceptualize nanotechnology was Richard P. Feynman, although he did not use this term in his speech at *American Physical Society* on December 29[th], 1959, where he made the first comments related to the subject. The word "nanotechnology" was first used by Professor Norio Taniguchi (1974) to define the fabrication at a scale of 1 nm. Nanotechnology is the potential ability to create anything from the smallest element, using the techniques and tools that are being developed today to place every atom and molecule in place. The use of nanometer implies the existence of a system of molecular engineering, which will likely generate the subversion of the factory manufacturing model as it is known.

The idea that nanotechnology has just a technological impact by itself in human activities and life and health must be analyzed, considering the strong impact on the economic activities in which the technologies are applied. In biomedical area for example, the considerable impact on pharmaceutical and biomedical companies is quite notorious.

Considering the whole discussion involved, it is pertinent to consider that rather important ethical questions are raised in this field. This discussion is important to define some borders, in view of the human and technical limitations, for the indescribable context of the problem. In this perspective of the discussion of the ethical issues, human values must not be neglected. In the nanotechnology area, particularly in the biomedical field, some questions are posed primarily on the impact on the human values and human condition, risks for human health derived from neglecting human behavior, for example, and on the borders of science. The possible violation of the basic principles of natural life and individuals' position concerning life borders, health and life, and its contextualization may be well understood.

In this work, some ideas are presented on nanotechnology, specifically about the concept and the reporting ideas involving nanotechnology. The discussion about the concepts is not the main objective of the present study. It is to show the importance of this technology, some of its applications and the exposure to different borders and significant risks in several areas behind the enormous fields of opportunities and challenges. Those risks may represent a problem but the challenges represent an open field to get significant progress.

The work is composed of the following parts:

1. Nanotechnology- a brief approach.

2. Nanopanels – a success of nanotechnology in industry.

3. Nanoelectronics – improving the life standard.

4. Nanotechnology in medicine-friendly efficient healthcare.

5. Ethics and nanotechnology.

6. Concluding remarks.

In section 1, nanotechnology is discussed in a general way. In section 2 nanotechnology and the applications in industry area, then in the electronics area (section 3), and in the medicine area (section 4) are presented and some discussion is proposed in order to define the boundaries for the advances on these areas. In the following section, nanotechnology is discussed in terms of ethics and in terms of the borders that nanotechnology applications must satisfy. In the last section, some final and concluding notes are presented, highlighting the results of the analysis. Important considerations are made about the close connection between ethics and the nanotechnology and the effects over the society and values. Finally, some future directions for the research are suggested. Some of this work is in (Ferreira *et al*, 2017).

## NANOTECHNOLOGY- A BRIEF APPROACH

In the $20^{th}$ century fifties, the discovery of the double helix structure of the deoxyribonucleic acid (DNA) molecule brought new perspectives to medicine and new hopes to mankind. The revolutionary discovery of DNA allowed advances on new fields of science. Since then, research brought many other significant discoveries which borders are completely unknown.

The $20^{th}$ century saw huge changes in the face of Earth with so many and so great discoveries that were totally impossible before. From the innovations that made possible to fly using heavy aircrafts, to use antibiotics for medical care, the arrival and enormous development of electronics, or whatever many other discoveries that revolutionized totally the human's life. The existence got a completely different sense. Newly, the introduction of the computer and the advances in electronics with alternating current circuits have contributed to life at another level of existence. The radio, television, cars, telephone, airplane, computers and internet services were products that have become available to the current life.

As can be easily seen and perceived every day, science has been facing important advances in many fields in the last decades and recently great new advances have been made. Now a new history may be in progression in the current century. Technology at a nanoscale has arrived and a completely new world is already there, with significant and continuous discoveries.

Feynman, cited in Freitas (2005), proposed the use of machine tools to make smaller machine tools, these to be used in turn to make still smaller machine tools, and so on all the way down to the atomic level. Feynman was clearly aware of the potential medical applications of the new technology he was proposing.

These smaller machine tools would come to be transformed in nanomachine tools, nanodevices, and nanorobots, which could ultimately be used to develop a wide range of atomically precise microscopic instrumentation and manufacturing tools, that is, nanotechnology.

This new technology, nanotechnology, which is an expression that comes from the Greek word "nano" ("dwarf", in English), is applied on a molecular level in engineering or manufacturing, for example.

The most common definition of nanotechnology is that of manipulation, observation and measurement at a scale of less than 100 nanometer (one nanometer is one billionth of a meter).

The term nanotechnology refers to several distinct classes of technology, each one with its own set of capabilities, potential applications and risks. The terms in this field that are specifically used for the different technologies may vary. However, it is important to be aware of the fundamental meaning of each one and the important dissimilarities between them.

Besides, nanotechnology is inherently multidisciplinary, depending on analytical techniques and methodologies of a set of disciplines including chemistry, physics, electrical engineering, material science and molecular biology, worked on the basis of many scientific fields in the area of lithography, nanomachines or nanorobots, for example.

Nanotechnology makes possible the modification of individual atoms and of molecules at a precise location, either chemically or physically and makes possible to develop devices which can scan and manipulate objects at near atomic scale (see Kubik *et al*, 2005). Dealing nanotechnology with materials, devices, and their applications, in areas such as engineered materials, electronics, computers, sensors, actuators, and machines, at the nano length-scale, the advances brought an enormous impact on those areas. Medicine, engineering of materials or electronics are nowadays confronted with progress but also with an ethical discussion. Considering these nano-meter length scales, numerous disciplines and new technologies are going together to develop new processes and combinations.

Since the nineties, the progresses were very relevant (see for example, Srivastava and Atluri, 2002). The discoveries of atomically precise materials are very significant, particularly the advances in:

1. medicine, for example, in the implantation of nano-robots.

2. the scanning probe and manipulation techniques to image and manipulate atomic and molecular configurations in real materials.

3. the conceptualization and demonstration of individual electronic and logic devices with atomic or molecular level materials.

4. the self-assembly of materials to be able to put together larger functional or integrated systems.

5. computational nanotechnology, *i.e.*, physics, and chemistry, based modeling and simulation of possible nanomaterials, devices, and applications.

6. agriculture (for pest control).

The possibilities are endless on these and many other fields.

One of the fields developed in nanotechnologies is the area of nanorobotics, which involves devices as nanobots, nanoids, nanites, nanomachines or nanomites. These concepts have been used to describe this kind of devices, today much under research and development. Nanorobotics refers to the nanotechnology engineering discipline of designing and building nanorobots, with devices that are ranging in size from 0.1 to 10 micrometers and are constructed of nanoscale or molecular components.

The invention of nanorobot hardware architecture for medical defense should provide the basis for advanced 'computational nanomechatronics: a pathway for control and manufacturing nanorobots' (Cavalcanti, 2009).

From the many nanorobots applications, it is very interesting to observe the following case. In 2002, in a paper published in Nature, researchers of New York University have presented a major step in building a robust, controllable machine from DNA, by demonstrating a robust sequence-dependent rotary DNA device operating in a four-step cycle. The authors (Yan *et al*, 2002) show that DNA strands control and fuel this device cycle by inducing the interconversion between two robust topological motifs, paranemic crossover (PX) DNA and its topoisomer JX DNA in which one strand end is rotated relative to the other by 180°. The authors say that they expect that a wide range of analogous yet distinct rotary devices can be created by changing the control strands and the device sequences to which they bind.

Constructed from synthetic DNA molecules, the device improves upon previously developed nano-scale DNA devices. The researchers say that the new device may help to build the foundation for the development of sophisticated machines at a molecular scale, ultimately evolving to the development of nano-robots that might someday build new molecules, computer circuits or fight infectious diseases (see Yan *et al*, 2002).

Controlled mechanical movement in molecular-scale devices has been realized in a variety of systems, by exploiting conformational changes triggered by changes in redox potential or temperature, reversible binding of small molecules or ions, or irradiation.

"The incorporation of such devices into arrays could in principle lead to complex structural states suitable for nanorobotic applications, provided that individual devices can be addressed separately. But because the triggers commonly used tend to act equally on all the devices that are present, they will need to be localized very tightly. This could be readily achieved with devices that are controlled individually by separate and device-specific reagents. A trigger mechanism that allows such specific control is the reversible binding of DNA strands, thereby 'fuelling' conformational changes in a DNA machine" (Yan *et al*, 2002).

This new robust sequence-dependent rotary DNA device operating in a four-step cycle represents an improvement upon the initial prototype system that used this mechanism but generated by-products. That initial prototype system was announced in 1999 by Prof. Seeman that consisted in a machine constructed from DNA molecules, which had two rigid arms that could rotated from fixed positions by adding a chemical to the solution. The problem was that this chemical affected all molecules within a structure uniformly. A big advance now is the fact that the movement can be manipulated within molecule pairs without affecting others within a larger structure, what is got by inserting DNA set and fuel strands into individual molecule pairs.

In biological research, particularly in the area of molecular biology and working nucleic acid molecules, many other achievements have been recently made. Moreover, many progresses have been made in the area of molecular biology applications. In this field, many techniques have been successfully implemented in many scientific fields, such as in the diagnostics area or in biology, biotechnology or medical science. For example, "the introduction of polymerase chain reaction (PCR) resulted in improving old and designing new laboratory devices for PCR amplification and analysis of amplified DNA fragments. In parallel to these efforts, the nature of DNA molecules and their construction have attracted many researchers. In addition, some studies concerning mimicking living systems, as well as developing and constructing artificial nanodevices, such as biomolecular sensors and artificial cells, have been conducted" (Kubik *et al*, 2005).

In this century, the development of ribonucleic acid (RNA) and deoxyribonucleic acid (DNA) diagnostics are being made considering a large set of approaches using new techniques and technologies, many of them associated to DNA developments, being particularly supported by nanotechnology solutions, which have been already successfully introduced. These achievements take part in the fundamental development of many areas of diagnostics and science, including the health care, medicine or pharmaceutical industry (Kubik *et al*, 2005).

There are great scientific revolutions in human history and as seen before, nanotechnology is certainly one of them, which opens up unimaginable possibilities in various fields of human reality and in various fields for application in the industry.

## NANOPANELS - A SUCCESS OF NANOTECHNOLOGY IN INDUSTRY

Nanotechnology has already been used in many industries. In this section some Brazilian cases are referred, taking into account Chavaglia and Filipe's recent research for solar panels in Brazilian market, introducing nanopanels technology. Some specific examples for Brazil are shown.

Considering the promising importance of nanopanels: in the future this technology may be applied to telemedicine with important consequence, for example. They will be dominant in this section. For instance, energy may be supplied to nano-biomedical data transmitters in future body area networks.

Nanotechnology will offer, in addition to products of higher quality at lower cost, a range of possibilities to produce new means of production and new types of resources and factors. This is a manufacturing system that can produce more manufacturing systems (plants that produce other plants) in a quick, cheap and clean way. The means of production may be reproduced exponentially. So, in just a few weeks, power would pass from a few to several billion nanofactories. Thus represents a kind of revolutionary technology, manufacturing, powerful, but also with many potential risks, in addition to the benefits it has (see Euroresidentes, 2011, p. 01).

It is estimated that only developed countries will allocate a sum of around USD 5.5 billion to this subject.

An important example of successful application of nanotechnology in Brazil is *Empresa Brasileira de Agropecuária* (Embrapa). It has been working with nanotechnology in various research centers and has already released some products. One of the most notable is perhaps the "electronic tongue", a device that combines chemical sensors with nanometer-thick with a computer program that detects flavors and aromas and serve to quality control and certification of wines, juices, coffees and other products. In Brazil, the budget of the Ministry of Science and Technology for the next four years is R 680 million.

The subversion of the industrial model is directly related to the use of nanotechnology in the various branches of economic activities. Follows an outline for the manufacture of electric power, which generically McKibben considered itself as a country's economy (see McKibben *et al*, 2009, p. 24).

In fact, the use of nanotechnology in the production processes of industry is a reality. It can be seen, for example, the particular case of the production of electric energy produced by photovoltaic panels. The emphasis is placed on the competitive advantage associated to the use of nanotechnology to solar energy production for companies in this market segment.

The American firm "Nanosolar", which studies are sponsored by major companies like Google or IBM or resulting from the allocation of benefits offered by the Department of Energy, is leading the race for energy production derived from nanotechnology. The company has named this technology as "nano-photovoltaic panels".

Note that to produce energy derived from solar radiation it is necessary to understand the functioning of a photovoltaic cell. The photovoltaic solar energy is obtained through direct conversion of light into electricity, the so-called "photovoltaic effect" (see Becquerel, 1839, the discoverer of this effect).

The photovoltaic cell works when light reaches the photovoltaic panel and moves electrons which circulate freely from atom to atom, forming the electric current.

The photovoltaic cell is a practical application of photoelectric effect. When the light falls on certain substances, takes off electrons that circulating freely from atom to atom, form a chain that can be stored. The photovoltaic cell that transforms the light into electricity will continue to generate power according to the level of radiation emitted, i.e. while the panel receive the light it will continue to generate electricity.

According to Nascimento (2004, p. 02) "the photovoltaic cell does not store electrical energy but maintains a flow of electrons established in an electrical circuit as long as there is light on it", as it is schematized in Figure 1.

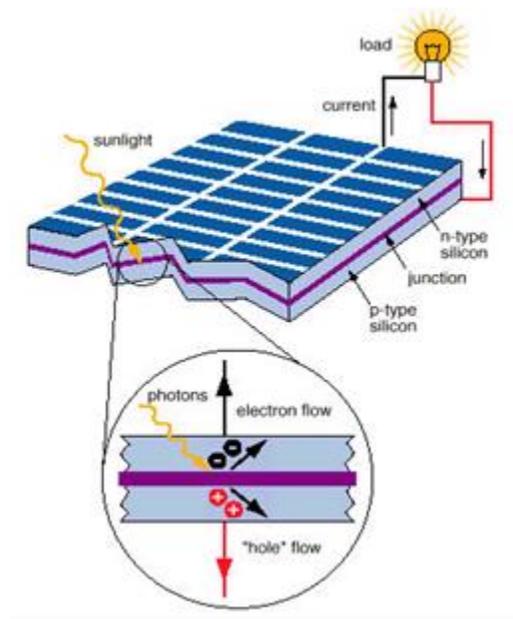

Figure 1- Energy Conversion: Direct conversion of solar radiation into electricity (Source: Australian CRC for Renewable Energy Ltd)

See below in Figure 2 the schema of a complete photovoltaic installation.

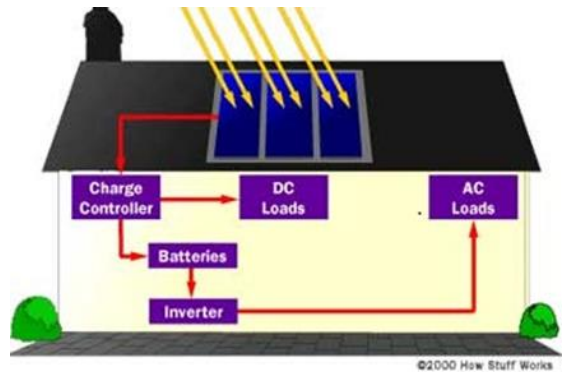

Figure 2 - Photovoltaic system installation: Schema of a complete photovoltaic installation. (Source: Toothman, J. and Aldous, S., 2000)

The use of this product has been showing a reduction in total costs on firms in relation to other types of solar energy and meet government requirements and social use of energy for clean sources. Indeed, note that the energy produced with the classic solar panels is much more expensive than the energy produced in either thermoelectric power plants or dams. If instead nano-photovoltaic panels are used, although, more expensive, the price is already comparable with the price of the energy produced in either thermoelectric power plants or dams (see Ferreira *et al*, 2014).

Companies need to be dynamic in the development of innovations and thus creating competitive advantages through their production processes, they can create economic value, and consequently generate their viability in the market in which they operate. For McDonough III (2009, p. 04) "in the current buoyant economy, organizations must continually reinvent what they are and what they do [...]". This means that they need to constantly maintain market differentiation, through deliberate strategies in order to obtain competitive advantages that provide monopoly profits, even if temporary in this environment, requiring from companies a high degree of competitiveness.

The competition is part of a dynamic and evolutionary operation of the capitalist economy. The evolution of this economy is seen as over time based on an uninterrupted process of introduction and diffusion of innovations in a broad sense. There are significant changes in the economic space in which these companies operate, whether they are changes in products, in processes, in sources of raw materials, in forms of productive organization, or in their own markets, including changes in terms of geography (see Schumpeter, Brazilian version, 1982, p. 65).

According to McAfee and Brynjolfsson (2008, p. 78) "the result is that an innovator with a better way of doing things can grow at unprecedented speeds and dominate the industry".

At the same time, it is interesting to see that nanotechnology is a technology that may use a minor quantity of natural resources than other technologies. Probably, a company gets additional benefits and has a lower cost and additionally gets better results for operating inputs. Surely, this has a consequent benefit for nature and environment, A higher level of competitiveness is got and there are better results also for society, as a whole. The companies will need to retain a smaller amount of the existing common resources for the optimization of their activities in the future and this will contribute probably for the preservation of resources. It is no longer possible to count on with such a supply of natural resources as that existing on last decades to meet the continuing huge demand, given the levels of production that humanity has achieved as standard over the last century.

To Nogami and Passos (1999, p. 03), from the harsh reality of scarcity arises the necessity of choice. Since it is not possible to produce everything that people want, mechanisms must be created to show somehow to the societies the path to decide what goods to produce and what needs have to be met.

The possibility of economic and technological efficiency reflects a producing combination of forces and inputs to reach an interesting production level for the company. All the means or methods of production indicate different combinations. Production methods vary according to the way such combinations occur. Every concrete act of production incorporates a kind of combination. A company can itself be considered a combination, and even may be the conditions of production of the whole economic system (see Schumpeter, 1982, p. 16).

Thus, it can be said that any company using clean energy in its production of goods and services (in particular, the use of photovoltaic panels generated by nanopanels) generates competitive advantage by breaking the closed circle of the economy. This is made by creating a new mechanism of generation of market value, since it is a new way to produce through a new combination of available resources.

In a direct way, besides the fact that the use of photovoltaics is already an innovation in itself, when it is combined with nanotechnology, its power of subversion, which can also be interpreted as its ability to break paradigms in the energy industry, takes the form of a powerful competitive weapon for the production units.

Given this context, nanotechnology combined with the production of solar energy becomes interesting for the production units, whether they are public or private.

In order to describe the way a nanopanel is manufactured, it can be said that first it is necessary to produce the semiconductor's nanoparticles (about 20mn in size, equivalent to 200 atoms in diameter). Then, aluminum sheets are placed in press, similar to those used in graphic paper. These aluminum sheets may be very dynamic in their use, because of their length and their width. This makes the product much more adaptable to formats required for the product. Then a thin layer of semiconducting ink is painted on the aluminum substrate. After that, another press put layers of cadmium sulfide and sulfur, and zinc oxide (CdS and ZnO). The layer of zinc oxide is non-reflective to ensure that sunlight is able to reach the semiconductor layer. Finally, the sheet is defined in sheets of solar cells. Unlike other methods of panels' manufacture that are usually used, which typically requires a special location for manufacturing, nano-panels can be produced outdoors (Ferreira *et al*, 2014).

This technology is very interesting, reaching good results either in terms of costs when compared with other solar panels and also in terms of performance even considering public energy. Those costs have coming to be considerably reduced.

In Brazil a strong trend of increasing costs to the production of hydroelectric power is observed. The comparison of the hydroelectric energy produced in Brazil and other countries like Canada, for example, shows the importance of the development of alternative energies. In Brazil, the behavior of the total cost of producing energy used to be well above inflation in the country for a long period. Recently, Brazil took the advantage of using photovoltaic energy using nano-panels. It may be noted in fact that this is one of the reasons that qualify the use of photovoltaic energy as a generator of competitive advantage in the market. The company which owns such technology may experience a reduction on its variable costs, compared with solar photovoltaic panels (silicon). Nano-panels allow also to perceive a competitive advantage in environmental conservation terms, when compared with the energy provided by public network (Ferreira *et al*, 2014).

Therefore, nanotechnology is contributing for the transformation of traditional models. This transformation may be in the way goods and services are produced, or in the way the production is conducted and made. Also in terms of nanomedicine, it is expectable that nanopanels may have interesting applications, for instance, in clothing or skin for energy supplying.

**NANOELECTRONICS-IMPROVING THE LIFE STANDARD**

What follows now may be seen as an example of how nanotechnology is contributing for the transformation of traditional models.

It is commonly accepted that electronic devices are essential in the support of nowadays life. And there is a great fundamental hope in nanotechnology to improve the capabilities of electronic components, namely in what concerns the reducing their dimensions, weight, and power consumption.

Indeed, being the electronic components traditionally of very short dimensions and very low weight, the reducing of these quantities to the nanoscale level makes possible to conceive electronic devices with much more qualities, since then it is possible to incorporate there much more components and even so to reduce the size and weight of the device.

As for the reducing of power consumption, only to emphasize that this is crucial in these times in which the energy shortage becomes an increasingly dramatic problem, nothing being too much in the efforts to achieve that the human activity happens to follow with an increasingly lower consumption of energy.

Also, to mention that the nanotechnology developments may also allow to conceive new kinds of electronic components, to be used either in classical electronic devices or modern ones.

All this to improve the humanity life standard. When speaking about the applications of nanotechnology to electronics it is usual to designate it as nanoelectronics, being the emergence of this word a sign of the huge importance of nanotechnology in electronics.

Nanoelectronics holds some answers for how it is possible to increase the capabilities of electronics devices and simultaneously reduce their weight and power consumption, making its operation more convenient and cheaper. Many of the nanoelectronics areas under development are devoted to the following topics:

-Improving display screens on electronics devices. This involves reducing power consumption while decreasing the weight and thickness of the screens.

-Increasing the density of memory chips. Researchers are developing a type of memory chip with a projected density of one terabyte of memory per square inch or greater.

-Reducing the size of transistors used in integrated circuits. One researcher believes it may be possible to "put the power of all of today's present computers in the palm of your hand".

Follow an enumeration of applications under development, falling in the topics described above (confer with the site in reference 50)), which brief description is enough auto-explicative:

1. Cadmium selenide nanocrystals deposited on plastic sheets have been shown to form **flexible electronic circuits**. Researchers are aiming for a combination of flexibility, a simple fabrication process and low power requirements.

2. Integrating **silicon nanophotonics** (confer with the former section) components into CMOS integrated circuits. This optical technique is intended to provide higher speed data transmission between integrated circuits than is possible with electrical signals.

3. Researchers at UC Berkeley have conceived a low power method to use **nanomagnets as switches**, like transistors, in electrical circuits. Their method might lead to electrical circuits with much lower power consumption than transistor based circuits.

4. Researchers at Georgia Tech, the University of Tokyo and Microsoft Research have developed a method to print prototype circuit boards using standard inkjet printers. **Silver nanoparticle ink** was used to form the conductive lines needed in circuit boards.

5. Researchers at Caltech have conceived a **laser that uses a nanopatterned silicon surface** that helps produce the light with much tighter frequency control than previously achieved. This may allow much higher data rates for information transmission over fiber optics.

6. Building **transistors from carbon nanotubes** to enable minimum transistor dimensions of a few nanometers and developing techniques to manufacture **integrated circuits built with nanotube transistors**.

7. Researchers at Stanford University have constructed a method to make functioning **integrated circuits using carbon nanotubes**. In order to make the circuit work they developed methods to remove metallic nanotubes, leaving only semiconducting nanotubes, as well as an algorithm to deal with misaligned nanotubes. The demonstration circuit they fabricated in the university labs contains 178 functioning transistors.

8. Developing a lead-free solder reliable enough for space missions and other high stress environments using **copper nanoparticles**.

9. Using electrodes made from **nanowires that would enable flat panel displays to be flexible** as well as thinner than current flat panel displays.

10. Using **semiconductor nanowires** to build transistors and integrated circuits.

11. Transistors built in **single atom thick graphene film** to enable very high-speed transistors.

12. Researchers have developed an interesting method of **forming PN junctions, a key component of transistors, in graphene**. They patterned the p and n regions in the substrate. When the graphene film was applied to the substrate electrons were either added or taken from the graphene, depending upon the doping of the substrate. The researchers believe that this method reduces the disruption of the graphene lattice that can occur with other methods.

13. Combining gold nanoparticles with organic molecules to create a transistor known as a NOMFET (**Nanoparticle Organic Memory Field-Effect Transistor**).

14. Using carbon nanotubes to direct electrons to illuminate pixels, resulting in a lightweight, millimeter thick **"nanoemmissive" display panel**.

15. Using quantum dots to replace the fluorescent dots used in current displays. **Displays using quantum dots** should be simpler to make than current displays as well as use less power.

16. Making integrated circuits with features that can be measured in

nanometers (nm), such as the process that allows the production of integrated circuits with **22 nm wide transistor gates**.

17. Using nanosized magnetic rings to make **Magnetoresistive Random Access Memory (MRAM)** which research has indicated may allow **memory density of 400 GB per square inch**.

18. Researchers have developed lower power, higher density method using nanoscale magnets called **magnetoelectric random access memor**y (MeRAM).

19. Developing **molecular-sized transistors** which may allow us to shrink the width of transistor gates to approximately one nm which will significantly increase transistor density in integrated circuits.

20. Using **self-aligning nanostructures to manufacture nanoscale integrated circuits**.

21. Using nanowires to build **transistors without p-n junctions**.

22. Using buckyballs to build **dense, low power memory devices**.

23. Using **magnetic quantum dots** in spintronic semiconductor devices. Spintronic devices are expected to be significantly higher density and lower power consumption because they measure the spin of electronics to determine a 1 or 0, rather than measuring groups of electronics as done in current semiconductor devices.

24. Using nanowires made of an alloy of iron and nickel to create dense memory devices. By applying a current magnetized sections along the length of the wire. As the magnetized sections move along the wire, the data is read by a stationary sensor. This method is called **race track memory.**

25. Using **silver nanowires embedded in a polymer** to make conductive layers that can flex, without damaging the conductor.

26. IMEC and Nantero are developing a **memory chip that uses carbon nanotubes**. This memory is labeled NRAM for Nanotube-Based Nonvolatile Random-Access Memory and is intended to be used in place of high density Flash memory chips.

27. Researcher have developed an organic **nanoglue** that forms a nanometer thick film between a computer chip and a heat sink. They report that using this nanoglue significantly increases the thermal conductance between the computer chip and the heat sink, which could help keep computer chips and other components cool.

28. Researchers at Georgia Tech, the University of Tokyo and Microsoft Research have developed a method to print prototype circuit boards using standard inkjet printers. **Silver nanoparticle ink** was used to form the conductive lines needed in circuit board.

# NANOTECHNOLOGY IN MEDICINE- FRIENDLY EFFICIENT HEALTHCARE

Nanotechnology is giving continuous steps towards new developments in research and applications. It will be expanded to many areas of life and sciences, particularly in medicine, where the diagnosis, drug deliveries' systems in the body or the treatment of diseases are particularly important.

Applications of nanotechnology for treatment, diagnosis, monitoring, and control of biological systems were recently referred to as "nanomedicine" by the Institutes of Health (see Moghimi *et al*, 2005).

The origin of the concept of nanomedicine is related to the idea that nanorobots and related machines could be designed, manufactured, and introduced into the human body to execute cellular repairs at the molecular level. Nowadays, nanomedicine has multiplied in many directions, understanding the idea that the ability to structure materials and devices at the molecular scale can bring enormous immediate benefits in the research and practice of medicine (see Freitas, 2005).

The use of instrumentation techniques inside the human body for 'medical nanorobotics for diabetes control', 'nanorobotics for brain aneurysm', 'nanorobots for treatment of patients with artery occlusion', 'nanorobots for laparoscopic cancer surgery', has been established and now only requires further industrial implementation and commercialization. Aspects such as integrating and using 'nanorobotic architecture for medical target identification' can effectively contribute to the advance of several medical issues, thus improving biomedical engineering. Upcoming and current available technologies should be used to achieve a fully functional 'medical nanorobot architecture based on nanobioelectronics' (Cavalcanti, 2009).

According to (Moghimi *et al*, 2005), the application of nanotechnology to medical activities, according to the idea inherent to the concept of nanomedicine, involves the identification of precise targets (cells and receptors) related to specific clinical conditions and the choice of the appropriate nanocarriers to achieve the required responses while minimizing the side effects. They say yet that "mononuclear phagocytes, dendritic cells, endothelial cells, and cancers (tumor cells, as well as tumor neovasculature) are key targets".

In (Moghimi *et al*, 2005) also it is shown how important are, nowadays, nanotechnology and nanoscience approaches to particle design and formulation. The authors highlight that they begin to expand the market for many drugs and are forming the basis for a highly profitable niche within the industry. Their article highlight rational approaches in design and surface engineering of nanoscale vehicles and entities for site-specific drug delivery and medical imaging after parenteral administration.

Among the very relevant fields of nanotechnology use one is the development of artificial tissues, organs and cells (which are much investigated for the replacement of defective or incorrectly functioning cells and organs). The implantation of encapsulated cells is being studied for the treatment of diabetes, liver failure, kidney failure and for the use of encapsulated genetically engineered cells for gene therapy. Artificial cells uses are very relevant for drug delivery and other applications in biotechnology, chemical engineering and medicine (see Kubik *et al*, 2005). New researches are being directed to discover new techniques which may contribute for the artificial growth of organs and tissues on nanopatterned scaffolds, aiming to obtain internal tissues implants. The studies about the creation of nanostructures that can interact with and replace natural biological materials give many opportunities and hope for medical treatments.

Some of the medical areas in which nanomaterials have successful utilization are precisely the medical diagnosis, the proper and efficient delivery of pharmaceuticals or the development of artificial cells.

As seen above, many advances in nanomedicine have been worked. Their use in valuable medical diagnostics or clinical therapeutics brings significant results. In Freitas (2005) many examples of applications are presented, for instance, single-virus detectors, tectodendrimers, nanoshells, fullerene-based pharmaceuticals, or immunoisolation, gated nanosieves.

Particularly, it is interesting to show the case of the advances on DNA sequencing and the consequent benefits. An ultrafast DNA sequencing has been made possible. Branton's team at Harvard University used an electric field to drive a variety of RNA and DNA polymers through the central nanopore of an a-hemolysin protein channel mounted in a lipid bilayer similar to the outer membrane of a living cell. Branton first showed that the nanopore could rapidly discriminate between pyrimidine and purine segments along a single RNA molecule and later demonstrated discrimination between DNA chains of similar length and composition differing only in base pair sequence (Freitas, 2005).

A colossal increase has been got recently in the understanding of the way basic biological processes happen at a molecular level (see for example the very particular case of the human genome sequence). The importance of these developments continues to grow very fast and so the focus on the essential molecular mechanisms underlying the normal functioning of cells, tissues and organisms themselves. Molecular medicine, by itself, exploits molecular and cellular biology advances to characterize how normal cellular processes fail or are subverted in disease. Nanotechnology has an important role in this matter.

According to Freitas, cited by Kubik (2005), there are three important molecular technologies:

- Nanoscale-structured materials and devices are promising for advanced diagnostics and biosensors, targeted drug delivery and smart drugs, and immunoisolation therapies.

- Biotechnology offers the benefits of molecular medicine via genomics, proteomics, and artificial engineered microbes.

- Molecular machine systems and medical nanorobots will allow instant pathogen diagnosis and extermination, chromosome replacement and individual cell surgery *in vivo*, and the efficient augmentation and improvement of natural physiological function.

Operating in the human body, nanorobots can monitor levels of different compounds and store information in an internal memory. The use of nanodevices may permit to reduce the intrusiveness, increasing the patient comfort and to give a greater fidelity on results, once the target tissue can be examined in its active state in the actual host environment.

Nanorobots may be used to rapidly examine a given tissue location, surveying its biochemistry, biomechanics, and histometric characteristics in greater detail (see Kubik *et al*, 2005). If, or when this happen, this will help in better disease diagnosing.

In general, nanotechnology works the engineering of molecularly precise structures and molecular machines; and nanomedicine makes the application of nanotechnology to medicine, considering also the use of medical nanorobotics. Medical nanorobots can offer targeted treatments to individual organs, tissues, cells and even intracellular components. They can get involved in biological processes at the molecular level.

It is clear at this moment that nanotechnology will continue to offer very effective solutions in many medical areas, allowing new treatments, giving more efficiency to the classic treatments, with much less suffering for the patients. Nanomedicine may even give answers at the level of the control of human aging. This world is a completely new open world, with very large potentialities at diverse scales and dimensions, very particularly in the area of telemedicine. It is very interesting to understand that the extent of telemedicine and telehealth success will depend on how well the heath care system exploits the capabilities of advanced information technology. This technology can extend the reach of medical facilities and resources, promoting efficiency, productivity, accuracy in clinical decision making, coordination and integration (Ackerman et *al*, 2002).

## ETHICS AND NANOTECHNOLOGY

Computers have brought the ability to type and made handwriting more and more superfluous, even obsolete. Mobiles and their miniaturization brought the ability to generally communicate anywhere to anybody, making more and more useless the traditional phones. Nanotechnology, because of its negligible dimension of devices, will spread the use of devices that now are nuisance. For instance, the pacemaker that now is used like a "last resource", maybe soon will be used in a much larger scale if it will be possible to design one at a nanoscale.

However ethical questions rise with the development of nanotechnology. What are the borders for this kind of progress? Until where we must go? What are the implications in a later phase of development of these technologies?

Global discussion has begun about this theme since the beginning of the discoveries about this subject. Legal, ethical and social implications are irreversible and the discussion is usually asked to be hard.

Vanessa Nurock, cited in Bensaude-Vincent (2010), questions the standard view of an ethics for nanotechnology. She argues that none of the current trends in the discipline of ethics would qualify for application to nanotechnology. Then considering that neurotechnology – a rapidly growing field at the intersection between nano and biotechnology – can affect moral capacities of the brain, she suggests that ethics itself may be affected by nanotechnology. And she leaves open the question of a co-construction of ethics and bionanotechnology.

Freitas (2005) says that the society should be "able to muster the collective financial and moral courage to allow such extraordinarily powerful medicine to be deployed for human betterment, with due regard to essential ethical considerations".

At EU, there are some important concerns and discussion about the development of the integration of human beings and artificial (software/hardware) entities. A funded project in this area (ETHICBOTS) was created to promote and to coordinate a multidisciplinary group of researchers into artificial intelligence, robotics, anthropology, moral philosophy, philosophy of science, psychology, and cognitive science, with the common purpose of identifying and analyzing techno-ethical issues concerned precisely with the integration of human beings and artificial (software/hardware) entities.

Next generations, in the perspective of the utilization of nanomaterials, will have many benefits in all these chapters (health, economic, environmental, etc). By using this kind of advanced technology, with many more applications and much friendlier using, the conditions this technology offers will be very environmental friendly as well, as already seen.

Actually, there are irreversible and long-term impact consequences for future generations and for the environment. However, there are risks, as well. But in truth there are incredible potential advantages. And the discussion is all over the place. The borders are, of course, evident and there has to be an achieved balance between the path found to the new dimension of knowledge and the contours the problem has. The rub of the question is exactly in what measure criticisms may have into account the potential of the technology.

So, despite the high feasibility for the economy and the environment, there are some considerations regarding the ethical, human dignity and moral borders on nanotechnology that should be taken into account.

It is interesting to go to the definition of human dignity and its relationship to moral made in the "opinion of the European Group on ethics in science and new technologies to the European Commission", 2005, when discussing the ethical aspects of information and communication technologies (ICT) implants in the human body. The group makes allusion to the EU's draft Treaty that Establishes a Constitution for Europe, stating that "*Human dignity is* *inviolable. It must be respected and protected"* (Article II-61), and goes on to explain that *"the dignity of the human person is not only a fundamental right in itself but constitutes the real basis of fundamental rights" (*Declaration concerning the explanations relating to the Charter of Fundamental Rights). This group says yet that this explanation does not strictly define human dignity and that many writers have attempted to fill this gap. One such attempt suggests that human dignity is defined as follows: *"the exalted moral status which every being of human origin uniquely possesses. Human dignity is a given reality, intrinsic to human substance, and not contingent upon any functional capacities which vary in degree. (...) The possession of human dignity carries certain immutable moral obligations. These include, concerning the treatment of all other human beings, the duty to preserve life, liberty, and the security of persons, and concerning animals and nature, responsibilities of stewardship."*

The introduction of new devices to go further above the natural capabilities of the human being brings lots of concerns to scientific community. Desires, challenges but also concerns are always present… What are the consequences? To human beings, to life it, to other living beings, to environment, to biodiversity, to progress, to the civilization as a whole...

Which are the consequences about the limit for the human being as human? Are humans creating a new being? Are we prepared to get different, biologically and to become another thing? Are humans creating further "nano-digital devices" between humans that have access to these advanced technologies and humans that have not access? This discussion is recognized but probably the ethical discussion has not been enough promoted.

In the long term, which are the consequences that nanotechnology has in the environment? It is possible to predict that less damage in nature can be done if nanomaterials are considered instead large equipments are used. Besides, once the need of inputs to produce these devices is lesser, probably there is an additional advantage on this perspective for nature preservation. But what is the reverse of the medal?

Really, the implications and directions of nanotechnology need further discussion. The consequences of avoiding the discussion may be severe for public, considering rejection or fear of the effects. Is there a need for a regulatory authority? Which kind of mechanisms are needed to interfere in the nanotechnology's developments area?

In truth, it may be possible in general to consider 5 topics for discussion (see Mnyusiwalla *et al*, 2003) about ethics in this subject. They are equity, privacy, security, environment and metaphysical when discussing the relationship human-machine. About each one of them some considerations may be made. So, assuming these topics, it may be said the following, without going until the limit of the analysis:

- **Equity**: the classical problem of developed and developing countries may assure a new topic for discussion. In fact, technology and development are closely related. Some important subjects that developing countries are worried with are poverty reduction, the problems of energy, water and health and somehow biodiversity as well. Nanotechnology would help these countries to better conditions of population health and way of life. But it is costly to implement any project in this area and as often there are no financial resources to do that, consequently these countries will be the less benefited with these advances. The main problem happens with the access to the benefits of technology considering the poorest and the richest people in any country. Who may reach the advances of science and benefit from them? Anyway, it is true that poor, whatever they are, may benefit from these advances. It may depend on the will of politicians and on the investments of some of these countries, for example. A consequence would be lower needs for energy and a cleaner energy production as well as other environmental benefits, a safer drug delivery, the improvement on health through for example better prevention, diagnosis and treatment.

- **Privacy**: at this level, it is possible to consider the enormous innovations which can be reached, by improving surveillance devices or other devices that restrict human privacy. An individual may be facing situations of "invisible" microphones and cameras. This is a real problem if privacy is intended to be assured. Wireless monitoring provides healthcare providers and patients to be mobile and being able to exchange data when needed. But there is the reverse of the medal: privacy violation? It is true that wireless communication and new physiological sensors bring enormous implications in e-health. One point needed is about the necessary discussion about privacy and security.

- **Security**: certainly new powered weapons will be available very soon, as far as "invisible" microphones and cameras will be available. Will this ensure new ways of security or will be added to a potential "arsenal of bio-terrorism and techno-terrorism or even nano-terrorism?" (see Mnyusiwalla *et al*, 2003). Mnyusiwalla *et al*, 2003 also ask "who will regulate the direction of research in defensive and offensive military NT [nanotechnology]? How much transparency will be necessary in government and private NT initiatives to avoid misuses? There are also very interesting legal questions involving monitoring, ownership, and control of invisible objects".

- **Environment**: new materials (fullerenes, carbon nanotubes) are now available. No one knows what the precise effects will be when this kind of nanomaterials goes into the environment (and also the effects to health). What are the dangers? Where are they put in the end of their mechanical lives? Will some medical devices be kept inside human body? What are the consequences? Or will be given another destiny to them? Is possible to predict any kind of effects caused to the environment? Jacobstein (2006) refers that it is important to make an analysis on the risks associated to passive compounds in the less than 100 nanometer size range. There is the possibility, for example, of being introduced inadvertently in human bodies. There is in fact the concern with their ability to be inhaled, absorbed through the skin, or to pass through biological compartment barriers such as the blood brain barrier. This kind of dangers pose a real range of potential health and environmental risks that are associated to their potential toxicity or mutagenicity in their interactions with biological systems. While the range of effects vary, most of the risks may be addressed by advanced industrial hygiene and environmental health practices and techniques that seek to characterize the specific risks, exposure patterns, and control methods and enforce them through a combination of practitioner education, industry self-regulation, monitoring and government regulation. This is an important emerging field in the environmental and health sciences, since most of the existing legislation on environmental, safety, and health risks may cover particulates, but do not take the change in physical and biological properties at the nanoscale into account. It is reasonable to assume that passive nanoscale particle risks, although potentially serious if not addressed, will be characterized and addressed systematically under new versions or extensions to existing occupational, industrial hygiene, environmental, and medical regulations.

- **Metaphysical**: the incorporation of implants in human body, this is, the incorporation of artificial materials or machines into human systems may conduct to a concern around the human values and principles; will a human being be more than a human being? The borders are unknown. What kind of being and what kind of consequences. How this process may be controlled and accessed? One of the main issues is the "autonomy" of nano-bots, if they are able or not to move and to take decisions autonomously. Are they able to communicate each other and with the external environment? How can human beings ensure to "retire" them at the end of their "missions"? This is a very demanding question. The modification made on human bodies and living systems may create new realities. This is already a very real concern on the society, being in general the society often very skeptic about developments in this area. Catherine Larrère, cited in Bensaude-Vincent (2010), discusses "the recent trends in nanoethics which anchor ethics in metaphysics or theology by emphasizing the emergence of new relations of men to nature and to God. It is defended that the moral issue raised by the project of enhancing human performances does not really lie in going beyond the boundary of human knowledge and condition. It is more a question of the moral choice underlying this new form of hubris".

Nano-bots, or in general the "convergence" of bio-nano-info-neuro technologies, requires a new contract between science and society; society should not be seen as an afterthought only for making technology assessments, for social acceptability, etc.. However, they should be really involved in the process; "converging technologies" - like syn-bio - should not be developed "inside the walls" of research labs only by highly specialized scientists, such challenges should be introduced requiring a cross-disciplinary approach, where project teams include anthropologists, sociologist, philosophers, and ethicists ... with a "precautionary principle" in mind...

Researchers in ethicbots project in EU, which was finished in 2008, worked in order to identify techno-ethical case-studies on the basis of a state-of-the-art survey in emerging technologies for the integration of human and artificial entities; to identify and analyze techno-ethical issues concerned with the integration of human beings and artificial entities, by means of case-studies analysis; to establish a techno-ethically aware community of researchers, by promoting workshops, dissemination, training activities, as well as by the construction of an internet knowledge-base, on the subject of techno-ethical issues emerging from current investigations on the interaction between biological and artificial (software/hardware) entities; to generate inputs to EU for techno-ethical monitoring, warning, and opinion generation (see Capurro *et al*, 2005-2008).

Rodotà & Capurro (2005) show the limitations on ICT implants in the human body as deriving from an analysis of the principles contained in various legal instruments. They say that they should be assessed further by having regard to general principles and rules concerning the autonomy of individuals, which takes the shape of freedom to choose how to use one's body, *"I am the ruler of my own body"*, freedom of choice as regards one's health, freedom from external controls and influence.

## CONCLUDING REMARKS

The use of nanomaterials brought many benefits to mankind. Considering that nanotechnology is the ability to work at the atomic, molecular and supramolecular levels (on a scale of approximately 1 – 100 nm), the enormous possibilities to work nanomaterials can be easily understood. The creation and the use of these material structures, devices and systems with fundamentally new properties and functions result from their small structure, with applications in several areas, such as bioprocessing in industries; molecular medicine (considering physical, chemical, biological and medical techniques used to describe molecular structures and mechanisms, often referred as personalized medicine); analyzing the health effect of nanostructures in the environment; improving food and agricultural systems; or improving human performance at many fields, considering the different branches of science and the applications at many dimensions and scales.

Nanotechnology supplies the tools and the technology allowing that researches lead to the transformation of biological systems, as far as biology provides models and bio-assembled components to nanotechnology.

Nanoscales, used in biosystems, contribute to enhance very innovative and promising results in medical area. Improvements in the telemedicine and on health may be expected with new systems operation and new nanotechniques.

However, ethical as far as legal and social implications are posed. This discussion is as old as the issue is itself but the boundaries are very large, large enough to find out the difficulties on finding definitive answers to this discussion.

Ethical, legal and social implications in this context show the importance of nanotechnology to the society and the consequences that the development of nanotechnology will have to mankind in the future.

The irreversibility and long-term impact of these new developments enforce to have on mind that it is necessary to take into account also the rights of future generations and the planet (like Hans Jonas' "Principle of responsibility").

In fact, the future is there and came loudly. The innovations are going fast and although the endemic crisis, economic but also social and politic, will bring some restrictions to the investments in the area of health and research in many countries, new developments in this area will prevail and many discoveries will be made. Innovations are growing and development is there, as much as the will to go faster in many domains of scientific investigation. The research on this subject gets new results everyday and new challenges have to be faced. Some work is going on this way and it is intended, after the conclusion of this paper, to analyze some data in order to study the impact of the crisis on nanotechnology and on applications to medicine and industry. The crisis may bring some restrictions, but challenges have to be faced and a new beginning is there. Ethics concerning nanodevices and the borders of human beings considering nanotechnology in general and applied to medicine also need a new research to be developed next. This debate of the borders of the human being regarding the application of nanotechnology may be also enlarged in order to have a study of the impact into telemedicine and e-health.

17) Filipe, J. A., Chavaglia, J., Ferreira, M. A. M. (2015). A Note on Nano-Photovoltaic Panels Emergence in Energy Market. *International Journal of Latest Trends in Finance and Economic Sciences* 5(1), 841-845.

18) Filipe, J. A. (2015).Nanotechnology and Medicine Improvement. *International Journal of Academic Research* 7(2), 32-37. DOI: 10.7813/2075-4124.2015/7-2/A.5.

19) Floridi, L. (1999). *Philosophy and Computing: An Introduction*. London - New York: Routledge, 1999.

20) Floridi, L. (2007). A look into the Future Impact of ICT on Our Lives, *The Information Society* 23.1, 59-64, 2007.

21) Floridi, L. (2008). Information Ethics, its Nature and Scope in *Moral Philosophy and Information Technology*, 40-65, edited by J. van den Hoven and J. Weckert. Cambridge: Cambridge University Press.

22) Floridi, L. (2010). *Information - A Very Short Introduction*. Oxford: Oxford University Press.

23) Floridi, L. (2011). *The Philosophy of Information*. Oxford: Oxford University Press.

24) Freitas, R. A. (2005). What is nanomedicine?. *Nanomedicine: Nanotechnology, Biology and Medicine Vol.1, Issue 1*, 2-9.

25) Freitas, R. A. (2007). Personal choice in the coming era of nanomedicine. In Patrick Lin et al. (Eds.), *Nanoethics: The Ethical and Social Implications of Nanotechnology* (pp. 161-172). New York, NY: John Wiley.

26) Hristozov, D. & Malsch, I. (2009). Hazards and risks of engineered nanoparticles for the environment and human health. *Sustainability*, 1(4), 1161-1194.

27) Jacobstein, N. (2006). Foresight Guidelines for Responsible Nanotechnology Development. Draft version. Foresight Institute. In *http://www.foresight.org/guidelines/current.html#Intro.*

28) Lemos, R. T. S. & Kern, V. M. (2009). Technontologies, complexity, and hybrid interfaces. *TripleC*, 7, 29-37.
30